\documentclass{article}

\usepackage{PRIMEarxiv}

\usepackage[utf8]{inputenc} 
\usepackage[T1]{fontenc}    
\usepackage{hyperref}       
\usepackage{url}            
\usepackage{booktabs}       
\usepackage{amsfonts}       
\usepackage{nicefrac}       
\usepackage{microtype}      
\usepackage{lipsum}
\usepackage{fancyhdr}       
\usepackage{graphicx}       
\graphicspath{{media/}}     
\usepackage{cite}
\usepackage{amsmath}

\title{Image translation of Ultrasound to Pseudo Anatomical Display by CycleGAN}

\author{
  Lilach Barkat \\
  Technion Institute of Technology \\
  Haifa\\
  \texttt{lilachbarkat@campus.technion.ac.il} \\
  \And
  Moti Freiman \\
  Technion Institute of Technology \\
  Haifa\\
  \texttt{moti.freiman@bm.technion.ac.il} \\
  \And
  Haim Azhari \\
  Technion Institute of Technology \\
  Haifa\\
  \texttt{haim@bm.technion.ac.il} \\
}

\begin{document}
\maketitle

\begin{abstract}
Ultrasound is the second most used modality in medical imaging. It is cost effective, hazardless, portable and implemented routinely in numerous clinical procedures. Nonetheless, image quality is characterized by granulated appearance, poor SNR and speckle noise. Specific for malignant tumors, the margins are blurred and indistinct. Thus, there is a great need for improving ultrasound image quality. We hypothesize that this can be achieved, using neural networks, by translation into a more realistic display which mimics an anatomical cut through the tissue. In order to achieve this goal, the preferable approach would be to use a set of paired images. However, this is practically impossible in our case. Therefore, Cycle Generative Adversarial Network (CycleGAN) was used, in order to learn each domain properties separately and enforce cross domain cycle consistency. The two datasets which were used for training the model were “Breast Ultrasound Images” (BUSI) and a set of optic images of poultry breast tissue samples acquired at our lab. The generated pseudo anatomical images provide improved visual discrimination of the lesions with clearer border definition and pronounced contrast. Furthermore, the algorithm manages to overcome the acoustic shadows artifacts commonly appearing in ultrasonic images. In order to evaluate the preservation of the anatomical features, the lesions in the ultrasonic images and the generated pseudo anatomical images were both automatically segmented and compared. This comparison yielded median dice score of 0.91 for the benign tumors and 0.70 for the malignant ones. The median  lesion center error was 0.58\% and 3.27\% for the benign and malignancies respectively and the median area error index was 0.40\% and 4.34\% for the benign and malignancies respectively. In conclusion, these generated pseudo anatomical images, which are presented in a more intuitive way, enhance tissue anatomy and can potentially simplify the diagnosis and improve the clinical outcome.
\end{abstract}


\section{Introduction}
\label{sec:introduction}

Ultrasound (US) has become an essential part of our lives. US is probably the most cost-effective medical imaging modality available today. Stemming from its high availability and from the fact that it is considered hazardless, it has found applications in almost all fields of medicine. Even before birth we use it for monitoring fetus development. We use it routinely to detect and diagnose numerous diseases and pathologies. From cardiology to cancer and from gallbladder stones to fatty livers. We use it for guidance of invasive procedures and for laparoscopic surgery. Nonetheless, even though it was introduced to medicine about eighty years ago and despite the enormous progress in the fields of electronics and computers one factor remained unchanged: The displayed images are characterized by granulated appearance with poor SNR and substantial speckle noise. As a result, the anatomical visualization is relatively poor when compared to CT or MRI. Consequent to the above, the clinical information is compromised and image interpretation relies substantially on the skills of the practicing radiologist.

Attempts have been made to improved US imaging interpretability by implementing artificial intelligence (AI) and deep learning. Most of the research directions include speckled noise removal \cite{jeyalakshmi2010modified,rotman2019simultaneous}, image segmentation \cite{noble2006ultrasound} and tissue classification \cite{liu2019deep}. Deep learning has also found applications in ultrasonic imaging by introducing efficient and effective solutions for adaptive beamforming and adaptive spectral Doppler \cite{van2019deep}, and for accelerating image acquisition \cite{yoon2018deep,senouf2018high}. In addition, methods to improve US image visualization using CT as a reference have also been suggested \cite{vedula2017towards}. In recent years, there has been a significant development in the field of neural networks that sets the foundations for further improvement of ultrasound images using new approaches. 

The main stumbling block that hinders diagnosis in ultrasound imaging is probably the granulated and speckled appearance which requires analysis by an experienced radiologist \cite{smith2010introduction}. It is postulated here that a transformation into a more realistic and more intuitive display can be beneficial, by allowing a more straightforward interpretation. In this work we have chosen to transform the ultrasonic imaging into a pseudo anatomical display which seems to be more natural to comprehend. We hypothesized that this can be achieved using AI.

\section{Methods}
\label{sec:methods}
\subsection{Cycle Generative Adversarial Network (CycleGAN)}
In order to create a transformation between two image domains using AI, it is customary to establish a large number of paired images to train the network. This required that each pair of images would be acquired under the same exact conditions and views. Evidently, due to operator dependent manner of ultrasonic image acquisition, it may be impractical to obtain matching pairs of images that exactly correspond to the same anatomical cross sections. However, the introduction of Cycle Generative Adversarial Network (CycleGAN) \cite{zhu2017unpaired}, has laid the foundations for a new approach. CycleGAN, which has already been applied in several studies in the medical field  \cite{hiasa2018cross,jiao2020self,modanwal2020mri} is used to study the features of each image domain separately and enforce similarity in the cross domain. This, therefore, enables us to overcome the lack of paired images (i.e. ultrasound vs. optical/anatomical) datasets.

The CycleGAN architecture is comprised of two Generative Adversarial Network (GAN) units, one for each image domain. As shown in Fig.1, each of the GAN models has a generator and a discriminator which are trained simultaneously. The generator attempts to produce from the input image domain a realistic display which fits the target domain as good as possible. The discriminator on the other hand, examines the features of the produced image and decides whether or not it belongs to the target domain. According to the discriminator feedback, the generator tries to improve the generated image. In parallel, the discriminator accepts actual input images that belong to the target domain and improve its discrimination capability. Thus, the two networks enforce mutual improvement in an adversarial manner. 

In the context of this work, one generator was trained to produce pseudo anatomical display ($G_{PA}$) from ultrasonic images and its corresponding discriminator ($D_{PA}$) was trained to distinguish real from generated synthetic images as shown in Fig.1a. In an alternating fashion, the second generator was trained to produce US display ($G_{US}$) from optical anatomical images, and trained simultaneously its corresponding discriminator ($D_{US}$) (see Fig.1b). In addition, the two networks were used for examining the cycle consistency. More explicitly, cycle consistency assessment implies that the quality of transfer from ultrasonic images to anatomical display and back to ultrasonic images or from anatomical images to ultrasonic display and back was evaluated.

Referring to mathematical terms, the real ultrasonic image is defined as $x$ and the real pseudo anatomical image as $y$. Accordingly, the adversarial loss for each of the GAN unit is calculated by \cite{zhu2017unpaired}:

\begin{equation}
\begin{split}
    \mathcal{L}_{GAN_{PA}}(G_{PA},D_{PA}) = {} & \mathbb{E}_{y \sim pdata(y)}[log(D_{PA}(y))] + \\
                             & \quad  \mathbb{E}_{x \sim pdata(x)}[log(1 - D_{PA}(G_{PA}(x))] 
\end{split}
\end{equation}

\begin{equation}
\begin{split}
    \mathcal{L}_{GAN_{US}}(G_{US},D_{US}) = {} & \mathbb{E}_{y \sim pdata(y)}[log(D_{US}(y))] + \\
                             & \quad  \mathbb{E}_{x \sim pdata(x)}[log(1 - D_{US}(G_{US}(x))] 
\end{split}
\end{equation}

Where $L_{GAN_{PA}}$ and $L_{GAN_{US}}$ are the loss functions for US and PA respectively, and $E_{x \sim p data}$ and $E_{y \sim p data}$ are the corresponding expected values.

In order to enforce similarity in the cross domain, the input image was passed through one generator followed by the second generator, minimizing the discrepancy of the real input image from the reconstructed image. The cycle reconstruction loss $(L_{cycle})$ was calculated by:

\begin{equation}
\begin{split}
    \mathcal{L}_{cycle}(G_{PA},G_{US}) = {} & \mathbb{E}_{x \sim pdata(x)}[||G_{US}(G_{PA}(x))-x||_1] + \\
                         & \quad  \mathbb{E}_{y \sim pdata(y)}[||G_{PA}(G_{US}(y))-y||_1] 
\end{split}
\end{equation}

In the original paper of CycleGAN \cite{zhu2017unpaired}, the researchers proposed the usage of an identity loss to preserve the pixels color through the image translation. However, in our case, the two domains differ substantially in their color scale and contrast. In order to account for the opposite contrast between the two domains, i.e. enhancing ultrasound black colored masses translation into light colored optical images of the masses, the loss term was modified to take the negative value of the input images, i.e.

\begin{equation}
\begin{split}
    \mathcal{L}_{opposite}(G_{PA},G_{US}) = {} & \mathbb{E}_{x \sim pdata(x)}[||G_{US}(\bar{x})-x||_1] + \\
                              & \quad  \mathbb{E}_{y \sim pdata(y)}[||G_{PA}(\bar{y})-y||_1] 
\end{split}
\end{equation}

\begin{figure}[ht]
\centering
\includegraphics[width=0.5\paperwidth]{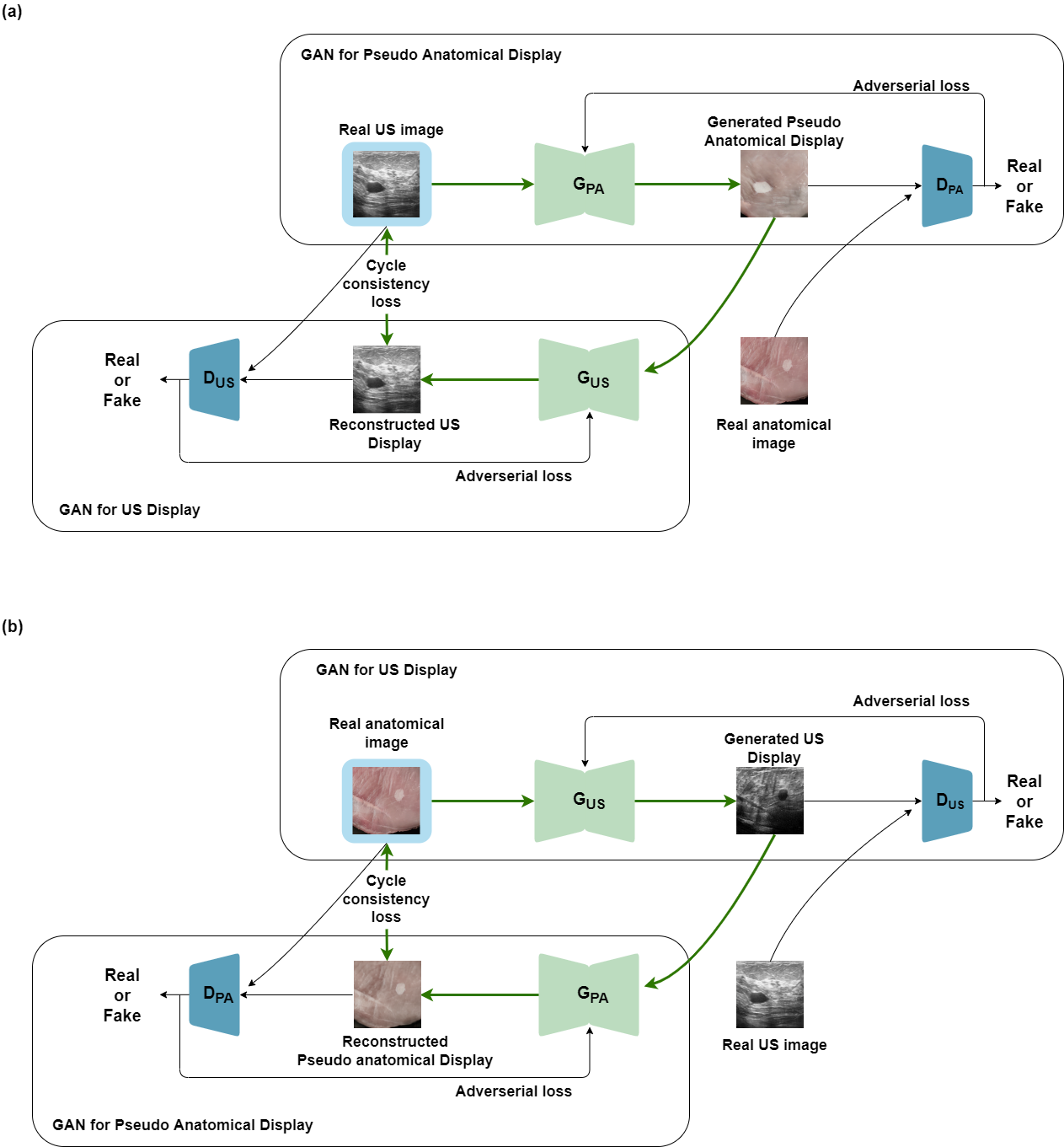}
\caption{(a) Schematic diagram of the CycleGAN model used here to translate real US images into pseudo anatomical images. The upper block is the GAN for producing pseudo anatomical display and the lower block is the GAN for producing the US display. (b) Identical network was used for generating US display from anatomical images.}
\end{figure}

\subsection{BUSI dataset}
A public dataset “Breast Ultrasound Images” (BUSI) collected from women in the age range between 25 and 75 years old \cite{al2020dataset} was used. The data was collected in 2018, including 600 female patients and consists of 780 images classified to three groups: (i) 133 normal images without masses (ii) 437 images with benign masses (iii) 210 images with malignant masses. The images were scanned by LOGIQ E9 ultrasound system and include additional manually traced masks of radiologist evaluation. The images are in a PNG format, vary in height and width and have an average size of 600x500 pixels. In our study, the data was pre-processed by removing text and label markings that are not part of the original ultrasound images and cropped to partially overlapping square patches of 450x450 pixels.

\subsection{Optic/Anatomic dataset}
To demonstrate feasibility, a set of optical images were collected from poultry breast tissue samples. To simulate abnormal tissue, small regions were thermally etched. The etched texture resembles the appearance of breast masses (see for example \cite{Radiology2013Assistant, samardar2002focal, franquet1993spiculated, sheppard2000tubular}). In order to simulate the typical US signal decay with depth, part of the images contained black background at the lower part of the image. The etched size, shape and location of the simulated tumors were created so as to resemble the relative distribution of the abnormal masses in the BUSI ultrasound images.

\subsection{Training parameters}
In order to train the model, the dataset, after being cropped to partially overlapping square patches, was divided into train, validation and test subsets. The training subgroup consisted of 80\% of the dataset images, 5\% for validation and the rest were used for testing the performance of the suggested method. The relative number of cases of malignant, benign and normal tissue were maintained in the validation and test groups as well. The network architecture was based on the CycleGAN architecture provided by \cite{zhu2017unpaired}. The hyperparameters were tuned according to the validation set, where the cycle loss was 10 times higher respectively to the GAN loss and the opposite loss was 0.03 lower. The model was trained on a NVIDIA Tesla V100 GPU running under Linux. Training computation time was approximately 36 hours.

\subsection{Automated segmentation}
In order to evaluate the quality of the generated images, image segmentation was performed and compared to the BUSI traced masks. The optical images were segmented by applying the Morphological Geodesic Active Contours (MorphGAC) \cite{marquez2013morphological}. As a preprocessing step to highlight the edges, the Inverse Gaussian Gradient (IGG) was applied. To control the steepness of the inversion, the alpha parameter was set to 100, and the standard deviation of the Gaussian filter sigma parameter was set to 1.5, for both the ultrasonic and the pseudo anatomical display images. 

Stemming from the fact that the BUSI tracing was preformed manually, it therefore inherently included inconsistencies. In order to overcome this problem, the generated images were also compared to automatically re-segmented masks by the MorphGAC algorithm for the ultrasound images. The code for the segmentation graphical user interface implementation is available \href{https://github.com/LilachBarkat/MorphGAC-Segmentation-GUI}{here}.

\subsection{Evaluation protocol}
Since image interpretability comparison is eventually subjective and evaluation metrics such as SSIM (Structural Similarity) which are commonly used in image reconstruction quality analysis are irrelevant in our case, other indices which are more indicative of the clinical merit were applied. The performance of the method was evaluated by comparing the optical segmented mask to both, the original BUSI reference masks, and to the BUSI re-segmented masks by the MorphGAC algorithm. Three metrices were used to evaluate the quality of the segmentation results. The first index used for contour evaluation was the Dice index which assesses shape similarity and is defined as \cite{dice1945measures}:

\begin{equation}
\begin{split}
    Dice = \frac{2\cdot TP}{(TP+FP)+(TP+FN)}
\end{split}
\end{equation}

where TP, FP, and FN are the true positive, false positive, and false negative pixels, respectively. Where positive means within the lesion mask and negative means outside the mask.

In addition, the accuracy in locating the center of mass of the lesion is defined as:

\begin{equation}
\begin{split}
    Center \: error = \\
     \frac{\sqrt{(C_{x_{Reference}}-C_{x_{Generated}})^2+ 
    (C_{y_{Reference}}-C_{y_{Generated}})^2}}{\sqrt{2}a}
\end{split}
\end{equation}

Where $C_{x_{Reference}}$, $C_{x_{Generated}}$ are the $x$ coordinates of the center of mass and $C_{y_{Reference}}$, $C_{y_{Generated}}$ are the $y$ coordinates of the reference masks, and the generated masks respectively. $\sqrt{2}a$ is a normalizing size factor corresponding to the diagonal length of the image (the largest possible dimension in the image). 

The Area index of the lesion is defined as:

\begin{equation}
\begin{split}
    Area \: index = \frac{|S_{Reference}-S_{Generated}|}{a^2}
\end{split}
\end{equation}

Where $S_{Reference}$, $S_{Generated}$ are the segmented lesion area of the reference and the generated masks, and $a^2$ is the size normalization to the image area.

\section{Results}
\label{sec:results}

A set of exemplary images translated from US to pseudo anatomical display is depicted in Fig.2. As can be observed, the pseudo anatomical display provides superior visual discrimination of the lesions with much clearer border definition and enhanced contrast. Furthermore, the tissue texture is more vivid and realistic. Importantly the algorithm manages to overcome the acoustic shadows artifacts commonly appearing in ultrasonic images, as can clearly observed in Fig.2.

\begin{figure}[ht]
\centering
\includegraphics[width=0.7\columnwidth]{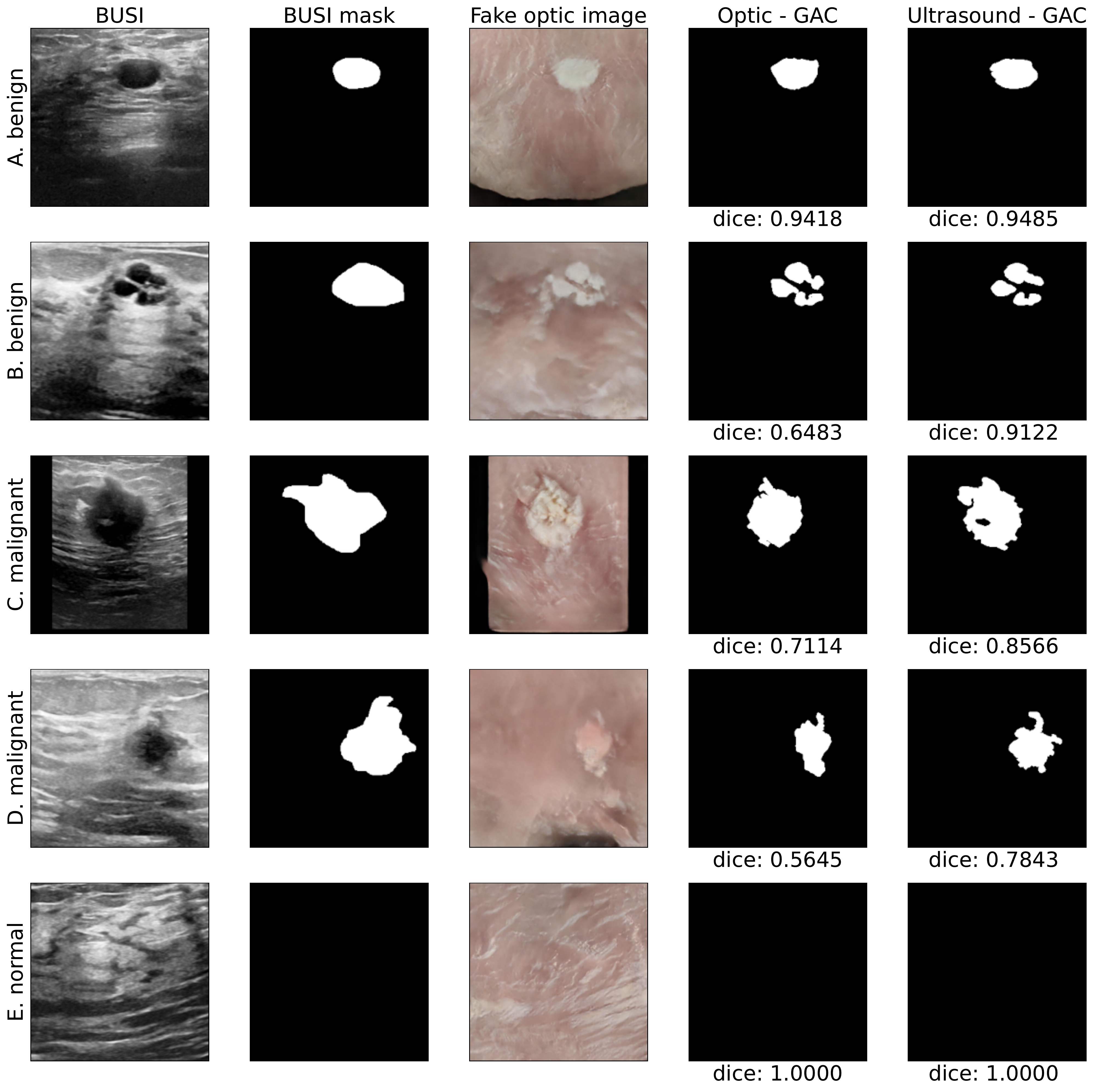}
\caption{(1st column) examples of the BUSI images. (2nd column) BUSI masks. (3rd column) the model generated pseudo anatomical images. (4th column) The corresponding segmented masks obtained by MorphGAC for the pseudo anatomical images and for the original ultrasound images (5th column), with the corresponding dice score. The rows correspond to: (A) Benign, (B) Benign, (C) Malignant, (D) Malignant and (E) Normal. As can be noted, the pseudo anatomical images are more natural to comprehend and the tumors are better defined.}
\end{figure}

\begin{figure}[ht]
\centering
\includegraphics[width=0.7\columnwidth]{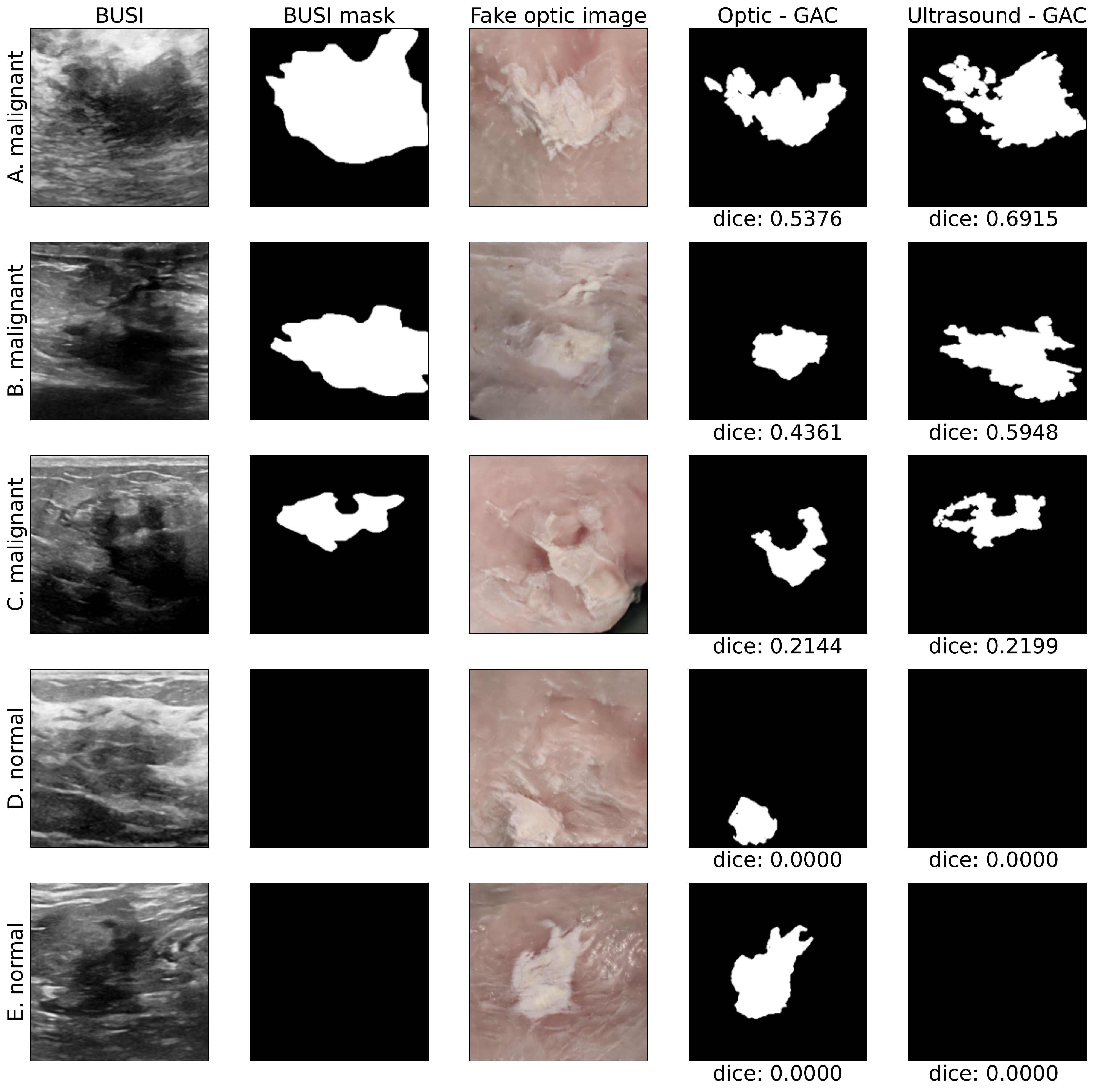}
\caption{(1st column) examples of the BUSI images. (2nd column) BUSI masks. (3rd column) the model generated pseudo anatomical images. (4th column) The corresponding segmented masks obtain by MorphGAC for the pseudo anatomical images and for the original ultrasound images (5th column), with the dice score. The rows correspond to: (A) Malignant, (B) Malignant, (C) Malignant, (D) Normal and (E) Normal. As can be noted, in these cases the algorithm was less effective.}
\end{figure}

Although the algorithm was successful in most cases, we have encountered several challenging cases in which the algorithm was ineffective. One typical problem was that it sometimes depicted lesions in normal tissues causing false positive readings (see for example Fig.3(d-e)). Another typical problem was generating lesion images which differed substantially from the reference BUSI masks (see Fig.3(a-c)). Nonetheless, part of the discrepancy can presumably be attributed to inconsistency in the BUSI tracing in “difficult to trace” ultrasonic images. This is demonstrated for example in Fig.4 where three substantially different tracings are presented for the same tumor. (example taken from ”malignant 3-9” in BUSI). As recalled, to overcome the variations stemming from the manual tracing, the ultrasonic images were re-segmented using the MorphGAC which yields better consistency. This is also demonstrated in Fig.4.

\begin{figure}
\centering
\includegraphics{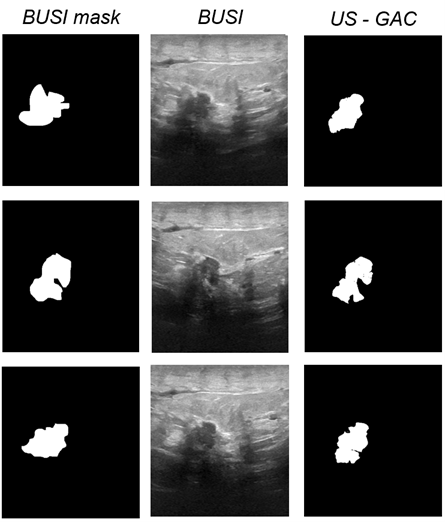}
\caption{(Middle column) Three exemplary original ultrasonic images of the same tumor which is marked as ”malignant 3-9” in BUSI. (First column) the corresponding three different tracings provided by BUSI for the same tumor. As can be observed, the three tracings differ substantially in their shape and geometry. (Last column). Contrary to that the corresponding MorphGAC segmented masks yielded more consistent tracings which appear to better match the lesion shape in the US images.}
\end{figure}

The overall estimation of the performance based on the criterions listed above, are outlined in Table1 and are also graphically depicted in Fig.5. As can be observed, the performance is better for benign tumors, this can be attributed to their more regular shapes. Contrary to that, for the malignant tumors which geometry is more irregular and which borders are commonly blurred, it was more difficult for the model and for the MorphGAC algorithm to accurately segment the lesions. 

\begin{figure}
\centering
\includegraphics[width=1\columnwidth]{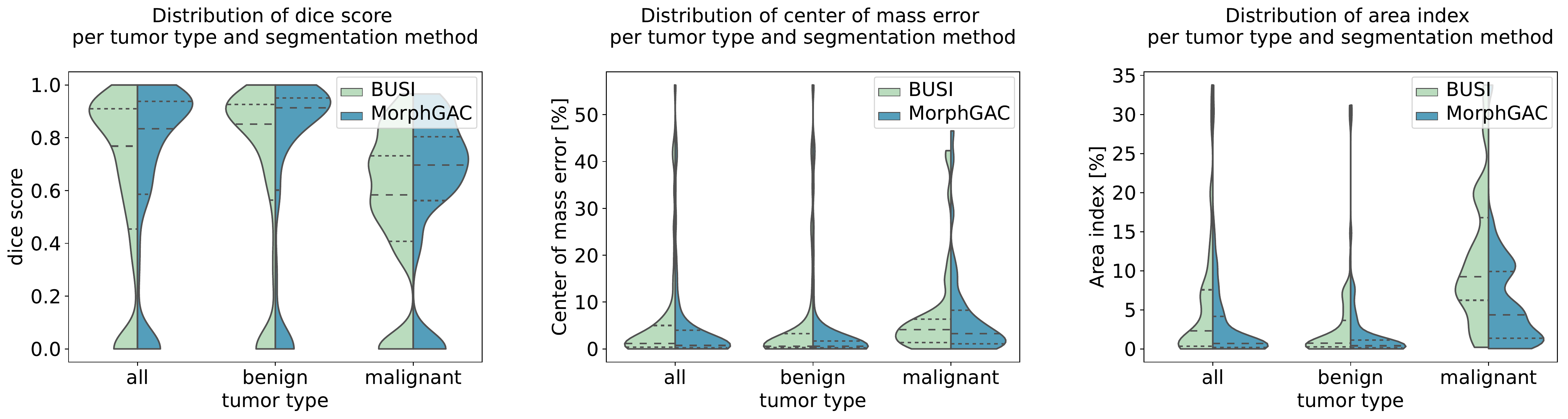} 
\caption{Distributions of the three metrices per tumor type for the two segmentation methods - BUSI (left) MorphGAC (right). (a) Dice distribution (b) Center error distribution (c) Area index distribution. The dashed lines represent the median score, upper and lower 25\% percentile.}
\end{figure}

Studying more closely the distribution of the quality indices, the area index and the center of mass error share similar behavior. For the benign tumor, both error indices are close to zero with a narrow distribution. Contrary to that, the distribution for the malignant tumors is wider. Although most of the masks for the generated images have a small detection error, some cases yielded significantly different masks. Examining the dice score distribution, for the benign tumors, the re-segmented masks have higher median score of 0.91 compared to 0.85 of the BUSI manually segmented masks. For the malignancies, the distribution is wider and the re-segmentation median score is 0.70 while for the manually segmented BUSI masks it is 0.58.

\begin{table*}
\caption{\raggedright Quantitative evaluation by comparing the optical segmented mask to both, the original BUSI reference masks, and to the BUSI re-segmented masks by the MorphGAC algorithm. \newline}
\centering
\begin{tabular}{llcccc}
\hline
\textbf{} & \textbf{Tumor type} & \multicolumn{1}{l}{\textbf{Median BUSI}} & \multicolumn{1}{l}{\textbf{Median MorphGAC}} & \multicolumn{1}{l}{\textbf{mean ± std BUSI}} & \multicolumn{1}{l}{\textbf{mean ± std MorphGAC}} \\ \hline
Dice & Benign & 0.85 & 0.91 & 0.67±0.36 & 0.70±0.38 \\
\textbf{} & Malignant & 0.58 & 0.70 & 0.53±0.30 & 0.60±0.32 \\ \hline
\textbf{} & all & 0.77 & 0.83 & 0.62±0.35 & 0.67±0.36 \\
 &  &  &  &  &  \\ \hline
Center error {[}\%{]} & Benign & 0.56 & 0.58 & 5.09±11.23 & 4.22±10.78 \\
 & Malignant & 4.13 & 3.27 & 7.21±10.29 & 7.21±10.65 \\ \hline
 & all & 1.17 & 0.73 & 5.76±10.95 & 5.14±10.79 \\
 &  &  &  &  &  \\ \hline
Area index {[}\%{]} & Benign & 0.74 & 0.40 & 2.84±5.21 & 2.11±5.08 \\
 & Malignant & 9.25 & 4.34 & 11.64±8.79 & 6.12±6.49 \\ \hline
 & all & 2.31 & 0.71 & 5.56±7.67 & 3.35±5.83
\end{tabular}
\end{table*}

\section{Conclusion}
\label{sec:conclusion}
In this paper, a method for image translation from ultrasound into pseudo anatomical display is presented. This image translation yields a more vivid and realistic tissue texture display which potentially can enable more straightforward comparison to the real anatomy. Furthermore, as can be observed, the pseudo anatomical images provide superior visual discrimination of the lesions. The borders are well delineated and the contrast clarity is substantially improved. This potentially can lead to faster and better diagnosis with improved clinical outcome.

The optical images used for training in this study were based on etched poultry breast tissue samples. Although the simulated lesions were prepared to resemble actual features of real breast tumors in terms of size and shape, optimal results presumably can be achieved by using actual optical images of anatomical cuts through breast tumor tissues. This however may require systematic photography during surgery, hence, it can be a great challenge to obtain.

\section{Acknowledgment}
\label{sec:acknowledgment}
The financial support of the Technion IIT, for Lilach Barkat scholarship is acknowledged.

\bibliographystyle{unsrt}  
\bibliography{main}

\end{document}